\documentclass[preprint,preprintnumbers,amsmath,amssymb]{revtex4}
\usepackage{booktabs}
\usepackage{mathrsfs}
\usepackage{epsfig}
\usepackage{graphicx}
\usepackage{dcolumn}
\usepackage{bm}
\usepackage{amsmath}

\let\jnfont=\rm
\def\NPB#1,{{\jnfont Nucl.\ Phys.\ B }{\bf #1},}
\def\PLB#1,{{\jnfont Phys.\ Lett.\ B }{\bf #1},}
\def\EPJC#1,{{\jnfont Eur.\ Phys.\ J.\ C }{\bf #1},}
\def\PRD#1,{{\jnfont Phys.\ Rev.\ D }{\bf #1},}
\def\PRL#1,{{\jnfont Phys.\ Rev.\ Lett.\ }{\bf #1},}
\def\MPLA#1,{{\jnfont Mod.\ Phys.\ Lett.\ A }{\bf #1},}
\def\JPG#1,{{\jnfont J.\ Phys.\ G}{\bf #1},}
\def\CTP#1,{{\jnfont Commun.\ Theor.\ Phys.\ }{\bf #1},}
\def\ZPC#1,{{\jnfont Z.\ Phys.\ C }{\bf #1},}
\def\JHEP#1,{{\jnfont JHEP \ }{\bf #1},}
\def\Rv{\not{\hbox{\kern-1pt $R$}}}
\def\p{\not{\hbox{\kern-3pt $p$}}}
\begin{document}

\title{Dark matter in the MSSM and its singlet extension}

\author{Jin Min Yang}

\address{
Institute of Theoretical Physics, Academia Sinica, Beijing 100190, China
\vspace*{1cm}}

\begin{abstract}
We briefly review the supersymetric explanation for the cosmic
dark matter. Although the neutralino in the minimal supersymmetric
model (MSSM), the next-to-minimal supersymmetric model (NMSSM) and
the nearly minimal supersymmetric model (nMSSM) can naturally
explain the dark matter relic density, the PAMELA result can
hardly be explained in these popular models. In the general
singlet extension of the MSSM, both the PAMELA result and the
relic density can be explained by the singlino-like neutralino.
Such singlino-like neutralinos annihilate into the singlet-like
Higgs bosons, which are light enough to decay dominantly to muons
or electrons, and the annihilation cross section can be greatly
enhanced by the Sommerfeld effect via exchanging a light CP-even
singlet-like Higgs boson. In this scenario, in order to meet the
stringent LEP constraints, the SM-like Higgs boson tends to decay
into the singlet Higgs pairs instead of $b\bar b$ and consequently
it will give a multi-muon signal $h_{SM}\to aa\to 4 \mu$ or
$h_{SM}\to hh\to 4 a \to 8 \mu$ at the LHC.
\end{abstract}
\maketitle

\section{Introduction}
The cosmic dark matter relic density measured by WMAP \cite{WMAP},
$0.0945 < \Omega h^2 < 0.1287$ , can be naturally explained by the
thermal production of WIMP (weakly interacting massive particle).
The neutralino (assumed to be the lightest supersymmetric
particle) in the minimal supersymmetric model (MSSM) is a good
candidate for the WIMP. Actually, the two events recently reported
by the CDMSII \cite{CDMSII} can also be naturally explained by
such a WIMP \cite{CDMSII-MSSM}. So both the the relic density
measured by WMAP and the two events observed by CDMSII can be
perfectly explained by the neutralino in the MSSM.

However, the excess of the cosmic ray positron in the energy range
10-100 GeV observed by PAMELA \cite{pamela} is hard to be explained
by the neutralino in the popular MSSM.  To explain the PAMELA
excess by WIMP annihilation, the WIMP must annihilate dominantly
into leptons since PAMELA has observed no excess of
anti-protons \cite{pamela} ( this statement may be not so solid due
to the significant astrophysical uncertainties associated with
their propagation \cite{kane}). Meanwhile, the WIMP annihilation
rate must be greatly enhanced (say by the Sommerfeld effect of a
new force \cite{sommerfeld2}) relative to the rate required by the
relic density if the dark matter is produced thermally in the
early universe. These two requirements cannot be satisfied in the
MSSM because there is not a new force in the neutralino dark
matter sector to induce the Sommerfeld enhancement and the
neutralino dark matter annihilates largely to final states
consisting of heavy quarks or gauge and/or Higgs
bosons \cite{jungman,neu1-4} (so if it predicts
positron excess, it must simultaneously gives antiproton excess).

Note that if supersymmetry is chosen by nature, the MSSM may not
be the most favored model to realize it. Actually, since the MSSM
suffers from the $\mu$-problem and the little hierarchy problem,
some non-minimal supersymmetric models may be equally or better
motivated, among which the most intensively studied is the
extension of the MSSM by introducing a singlet Higgs superfield.
If we do not impose any discrete symmetry to forbid some terms in
the superpotential, the model is the general singlet extension of
the MSSM. If we impose some discrete symmetry, then we obtain some
specified singlet extensions like the next-to-minimal
supersymmetric standard model
(NMSSM) \cite{NMSSM}
and the nearly minimal supersymmetric standard model
(nMSSM) \cite{xnMSSM}. As shown by recent
studies \cite{Hooper}, the nMSSM and NMSSM are unlikely to
explain the PAMELA result due to the tight parameter space
constrained by various current experiments while the general
singlet extension of the MSSM can perfectly make it. In the
general singlet extension of the MSSM, the singlino-like
neutralino (the lightest supersymmetric particle) serves as the
dark matter, which annihilates to the light singlet-like Higgs
bosons. Since the interaction between singlino-like neutralino and
singlet-like Higgs bosons is not suppressed and is typically the
weak interaction, the relic density can be naturally obtained just
like in the MSSM. At the same time, the singlet-like Higgs bosons,
not so related to electroweak symmetry breaking, can be light
enough to be kinematically chosen to decay dominantly into muons
or electrons. The Sommerfeld enhancement needed in the dark matter
annihilation for the explanation of PAMELA result can be induced
by exchanging the light CP-even singlet-like Higgs boson.

In this review, we recapitulate the recent studies on the dark matter explanation
in supersymmetric models. In Sec.\ref{sec2} we discuss the MSSM. In Sec.\ref{sec3}
we discuss the NMSSM and nMSSM. In Sec.\ref{sec4} we focus on the general singlet
extension of the MSSM.  Finally, a summery is given in Sec. \ref{sec5}.

\section{Neutralino dark matter  in MSSM}
\label{sec2}
The MSSM is the most economical realization of supersymmetry, which has
the minimal content of particles. Among the four neutralinos (the mixture
of neutral gauginos and neutral Higgsinos), the lightest one is usually assumed
to be the lightest superparticle (LSP).
Due to the R-parity conservation assumption, the neutralino
LSP is stable. Since it has weak interaction and its mass is around weak scale,
it is a perfect WIMP.
The neutralino LSP mainly annihilates
to final states consisting of heavy quarks or gauge and/or Higgs bosons,
as shown in Fig.1.

With current experimental constraints which are from the
precision electroweak measurements, the direct search for sparticles and Higgs
bosons, the stability of Higgs potential and the muon $g-2$ measurement,
a scan over the parameter space (see the last reference in \cite{NMSSM})
found that a large part of
the parameter space can give the required dark matter relic density. If we
project the allowed parameter space in the plane of $\tan\beta$ versus $\mu$,
it is shown in Fig.2.
\begin{figure}[htb]
\centerline{\psfig{file=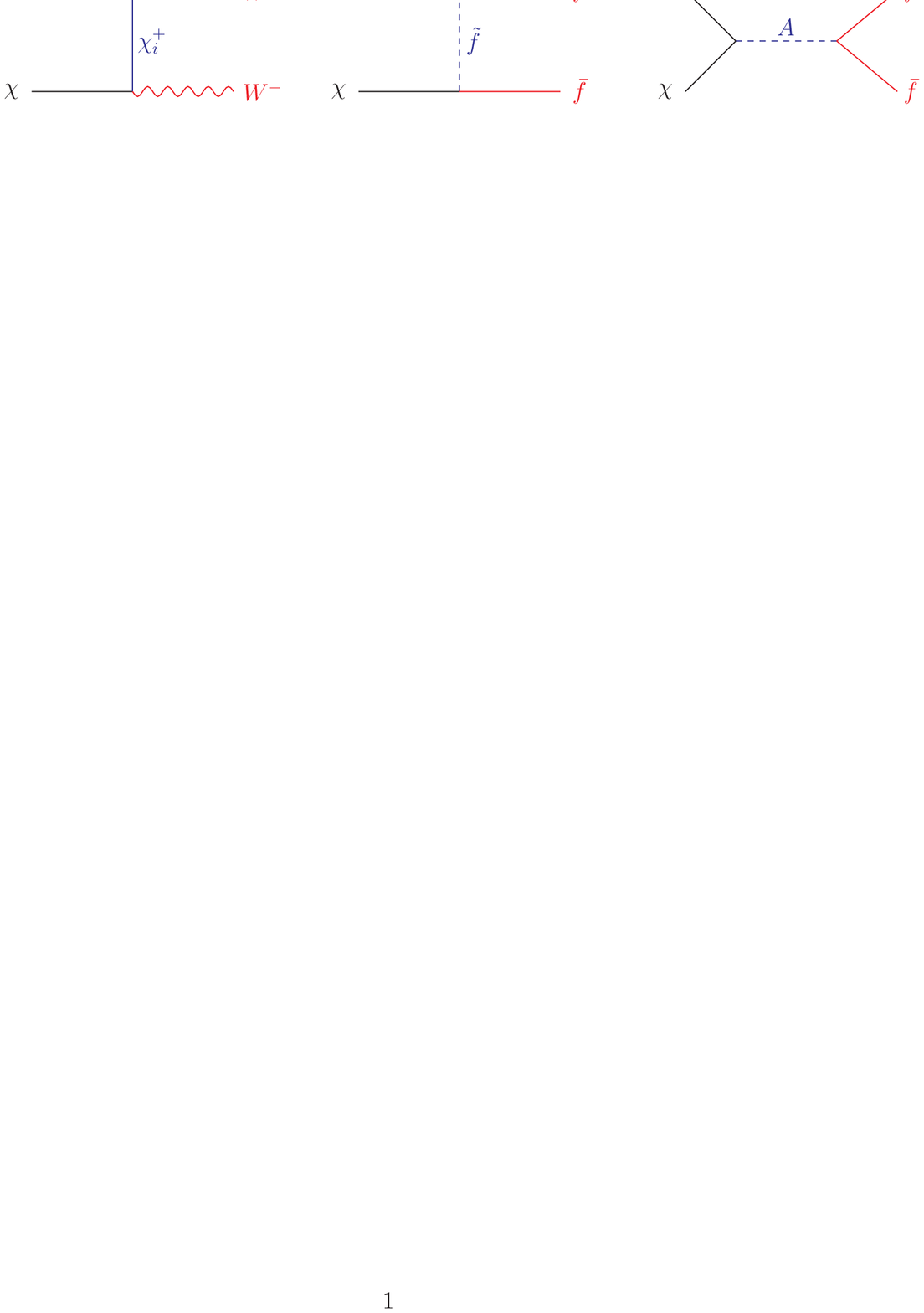,width=5.0in}}
\vspace*{-.5cm}
\caption{Feynman diagrams of the main annihilation channels of the neutralino LSP
         in the MSSM.}
\label{fig1}
\end{figure}
\begin{figure}[htb]
\centerline{\psfig{file=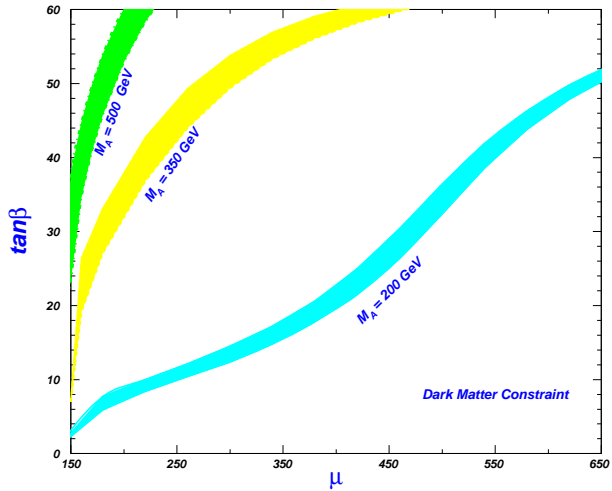,width=3.2in}}
\vspace*{-.5cm}
\caption{The shaded regions are allowed by the cosmic dark matter relic density
         at $2\sigma$ level plus other experimental constraints in the MSSM,
         taken from the last reference in \cite{NMSSM}.}
\label{fig2}
\end{figure}

In the constrained MSSM like mSUGRA, the parameter space allowed by the explanation
of dark matter relic density is usually displayed in the plane of $m_0$ versus $m_{1/2}$,
which showed that there exist several regions to give the required
dark matter relic density \cite{tata}.

Although both the neutralino LSP in the MSSM and the constrained
MSSM can naturally explain the dark matter relic density, the
PAMELA result cannot be explained. As shown in Fig.1, the
annihilation final states consist of heavy quarks or gauge bosons,
and, therefore, if it predicts positron excess, it must
simultaneously lead to antiproton excess. Meanwhile, there exist
no light scalar or gauge bosons to induce the Sommerfeld
enhancement.

\section{Neutralino dark matter in ${\rm NMSSM}$ and ${\rm nMSSM}$}
\label{sec3}
Both the  NMSSM and nMSSM extend the MSSM by adding a singlet Higgs superfield $\hat{S}$.
The difference of the two models is that the superpotential contains
a trilinear singlet term $\kappa \hat{S}^3$ in the NMSSM, which is replaced
by a tadpole term $\xi_F M_n^2 \hat{S}$ in the nMSSM.
There is no $\mu$ term in the superpotential and such a $\mu$ term is dynamically generated
through the coupling between the two Higgs doublets and the newly introduced singlet
Higgs field which develops a vacuum expectation value of the order
of the SUSY breaking scale. Thus the $\mu$ problem is solved in both models.
The little hierarchy problem can also be alleviated because on the one hand
the LEP II lower bound on the mass of the SM-like Higgs boson $h$ is relaxed by
the suppressed $ZZh$ coupling and/or by the suppressed visible decay
$h \to b\bar b$, on the other hand the tree-level upper bound on the Higgs boson mass $m_h$
is pushed up.
\begin{figure}[htb]
\centerline{\psfig{file=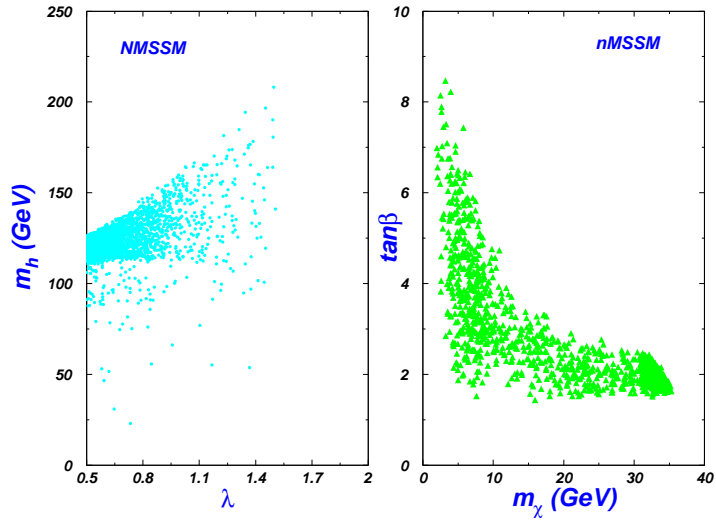,width=3.8in}}
\vspace*{-.5cm}
\caption{The shaded regions are allowed by the cosmic dark matter relic density
         at $2\sigma$ level plus other experimental constraints in the NMSSM and
         nMSSM, taken respectively from the last reference in \cite{NMSSM} and \cite{xnMSSM}.}
\label{fig3}
\end{figure}

In both models the neutralino LSP has a large component of singlino (the fermion component
of $\hat{S}$), which serves as the dark matter particle and can explain the dark matter
relic density measured by WMAP. With current experimental constraints which are from the
precision electroweak measurements, the direct search for sparticles and Higgs
bosons, the stability of Higgs potential and the muon $g-2$ measurement,
a scan over the parameter space was performed for the NMSSM
(see the last reference in \cite{NMSSM}))
and the nMSSM (see the last reference in \cite{xnMSSM}).
It was found that in both models there exist
a large part of the parameter space which can yield the required dark matter relic
density. In Fig.3 we show some results from the scan.

As shown in Fig.3, the neutralino LSP in the parameter space allowed by the
dark matter
relic density cannot explain the PAMELA result because in the NMSSM the lightest
CP-even Higgs boson cannot be light enough to induce the Sommerfeld enhancement
(the neutralino may explain either the relic density or PAMELA, but impossible
to explain both via Sommerfeld enhancement \cite{dm-nmssm})
while in the nMSSM the
neutralino mass is restrained in a narrow range.

\section{dark matter in general singlet extension of MSSM}
\label{sec4}
In the general singlet extension of the MSSM the Higgs superpotential contains
both the $\mu$ term and all possible terms of the singlet superfield. So this
model can only solve the little hierarchy problem but suffers from the $\mu$
problem. The dark matter candidate is the singlino-like neutralino LSP which
annihilates to the light singlet-like Higgs bosons $h$ (CP-even) or $a$ (CP-odd),
as shown in Fig.4,  and the relic density can be naturally obtained from the
weak interaction between singlino and singlet Higgs bosons.
Due to the vast parameter space, the singlet-like Higgs bosons
$h$ and $a$ can be light enough to be kinematically
restrained to decay dominantly into muons or electrons, as shown in the left
panel of Fig.5
The Sommerfeld enhancement can be induced by exchanging the light singlet
Higgs boson $h$ and for a light enough $h$ such an enhancement can be
quite large, as shown in the right panel of Fig.5.
\begin{figure}[htb]
\centerline{\psfig{file=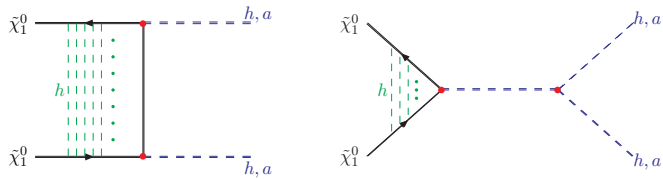,width=3.5in}}
\vspace*{-.5cm}
\caption{Feynman diagrams for the singlino-like neutralino dark matter
         annihilation where Sommerfeld
         enhancement is induced by exchanging a light CP-even Higgs boson $h$.}
\label{fig4}
\end{figure}
With the muon final states of the neutralino annihilation and the large
Sommerfeld enhancement induced by a light $h$, the PAMELA result can be
explained in this model \cite{Hooper}. A scan showed \cite{Hooper}
that in the allowed parameter space the SM-like Higgs boson $h_{SM}$
tends to decay into the singlet Higgs pairs $aa$ or $hh$ instead of
$b\bar b$. So the $h_{SM}$ produced at the LHC will give a multi-muon signal,
$h_{SM}\to aa\to 4 \mu$ or $h_{SM}\to hh\to 4 a \to 8 \mu$.
\begin{figure}[htb]
\centerline{\psfig{file=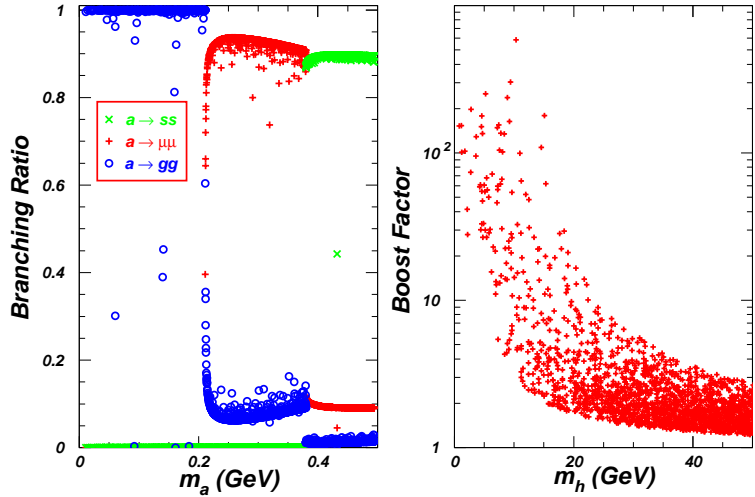,width=4.0in}}
\vspace*{-.5cm}
\caption{The left panel is the scatter plots showing the decay branching ratios
        $a\to\mu^+\mu^-$ (muon), $a\to gg$ (gluon) and $a\to s\bar s$ ($s$-quark).
        The right panel is the scatter plots  showing the Sommerfeld enhancement
        factor induced by $h$. These results are taken from the second reference
        in \cite{Hooper}.}
\label{fig5}
\end{figure}

\section{Conclusion}
\label{sec5} We briefly reviewed the supersymetric explanation for
the cosmic dark matter. The neutralino in the MSSM, NMSSM and
nMSSM can naturally explain the dark matter relic density, but it
can hardly explain the PAMELA result. In the general singlet
extension of the MSSM, both the PAMELA result and the relic
density can be explained simultaneously by the singlino-like
neutralino which annihilates into the singlet-like Higgs bosons.
These singlet-like Higgs bosons are light enough to decay
dominantly to muons or electrons, and the annihilation cross
section can be greatly enhanced by the Sommerfeld effect via
exchanging a light CP-even singlet-like Higgs boson. In this
scenario, in order to meet the stringent LEP constraints, the
SM-like Higgs boson tends to decay into the singlet Higgs pairs
instead of $b\bar b$ and consequently it will give a multi-muon
signal $h_{SM}\to aa\to 4 \mu$ or $h_{SM}\to hh\to 4 a \to 8 \mu$
at the LHC.

\section*{Acknowledgments}
This work was supported in part by the National Natural
Science Foundation of China under grant Nos. 10821504, 10725526 and 10635030.

\end{document}